\documentclass[aps,prb,showpacs,twocolumn,
preprintnumbers,amsmath,amssymb]{revtex4}
\usepackage{amssymb}
\usepackage{graphicx}
\usepackage{dcolumn}
\usepackage{verbatim}
\def\mc{\multicolumn}

\begin{document}

\title{First-principles study of the spin-lattice coulpling in spin frustrated
DyMn$_2$O$_5$}

\author{Tianqi Shen}
\affiliation{Key Laboratory of Quantum Information, University of
Science and Technology of China, Hefei, 230026, People's Republic of
China}

\author{Kun Cao}
\affiliation{Key Laboratory of Quantum Information, University of
Science and Technology of China, Hefei, 230026, People's Republic of
China}

\author{Guang-Can Guo}
\affiliation{Key Laboratory of Quantum Information, University of
Science and Technology of China, Hefei, 230026, People's Republic of
China}

\author{Lixin He
\footnote{corresponding author, Email address: helx@ustc.edu.cn} }
\affiliation{Key Laboratory of Quantum Information, University of
Science and Technology of China, Hefei, 230026, People's Republic of
China}

\date{\today}

\begin{abstract}

The lattice dynamic properties
and spin-phonon coupling
in DyMn$_2$O$_5$ are studied by
using the density-functional theory. The calculated phonon frequencies are in
very good agreement with experiments.
We then compare the phonon modes calculated from different
spin configurations.
The results show that the phonon frequencies
change substantially in different spin configurations, suggesting that the
spin-phonon coupling in this material is very strong.
Especially, the short range spin ordering has drastic effects on the
highest Raman and IR phonon modes that might be responsible for the
observed phonon anomalies near and above the magnetic phase transitions.

\end{abstract}
\maketitle


\section{Introduction}

Manganese oxides including RMnO$_3$, \cite{kimura03,goto04}
and RMn$_2$O$_5$ (R=Tb, Dy, Ho, etc.)
\cite{hur04,chapon04,blake05} are a very special class of improper
ferroelectrics that display strong 
magnetoelectric (ME) effects, such as 
the ``colossal magnetodielectric'' (CMD) effects
and magneto-polarization-flop effects. \cite{kimura03,goto04,hur04b}
These materials have attracted great attention,
\cite{aguilar06,cheong07,katsura05, sergienko06}
because the strong ME effects
can be utilized to
tune the electric properties of the materials 
via applied magnetic filed, or vice versa,
and therefore have
great potential for future multifunctional
device applications.
The {\it macroscopic} ME effects have their
{\it microscopic} origin
from the interplay between the lattice degree of freedom and spin degree
freedom.
Neutron scattering, \cite{chapon06}
as well as first-principles calculations \cite{wang07,wang08} shows that
the improper ferroelectricity in TbMn$_2$O$_5$ is driven
by the non-central symmetric magnetic ordering, which, by coupling to the
lattice, lower the crystal symmetry. \cite{chapon06,wang07}
The spin-lattice coupling can also
modify the lattice dynamitic properties.
Indeed, recent
temperature dependent Raman measurements \cite{flores06}
show anomalous phonon shift
at T$^*$ $\sim$ 1.5 T$_N$ and near the magnetic phase
transition temperature T$_N$ in DyMn$_2$O$_5$.
Similar anomalies are also observed for the IR modes. \cite{cao08b}
Further B-field dependent measurements show that the IR modes are very
sensitive to local magnetic structure, \cite{cao08}
which suggest that the spin-phonon coupling in this compound
is very strong.
However, the experiments give only the overall effects, and
the detailed mechanism of the spin-phonon coupling
is still lack.

First principles method has been applied successfully in studying
spin-phonon coupling in geometrically frustrated spinel ZnCr$_2$O$_4$.
\cite{fennie06}
It is a challenge problem to study the 
spin-phonon coupling in
the spin frustrated systems, such as RMn$_2$O$_5$, because
in these materials, the magnetic interactions
are of different magnitudes 
(e.g., $|J_4|$, $|J_5|$ $\gg$ $|J_3|$), \cite{chapon06,wang08} 
and the spin frustration leads to complicated
spin-spin correlations as functions of temperature
and magnetic field.
Therefore, the spin will develop different short range 
ordering at different temperature 
above the magnetic phase transition, and make it hard to identify the origin
of the phonon anomalies. \cite{flores06}
In this work, we study the spin-phonon coupling in DyMn$_2$O$_5$, to shed some
light on the observed phonon anomalies. We show that the short range spin
correlation and local spin structure have significant effects to the phonons
frequencies in this material.

\section{Methodology}

The first-principles
density-functional (DF) calculations are done by using
the Vienna \emph{ab initio} Simulations Package (VASP).
\cite{kresse93,kresse96}
We use a spin-polarized generalized gradient approximation (GGA),
with Perdew-Burke-Ernzerhof functional.\cite{perdew96}
A plane-wave basis and projector
augmented-wave (PAW) pseudopotentials \cite{blochl94} are used, with
Mn 3\emph{p}3\emph{d}4\emph{s}, and  Dy 5\emph{p}5\emph{d}6\emph{s}
electrons treated self-consistently.
A 500 eV plane-wave cutoff results in good convergence of the total
energies. We relax the structure until the changes of total energy
in the self-consistent calculations are less than 10$^{-7}$ eV, and
the remaining forces are less than 1 meV/\AA.
DyMn$_2$O$_5$ undergo several magnetic phase transitions \cite{blake05}
at low temperature.
We are interested in the phonon properties
in the paramagnetic (PM) and anti-ferromagnetic (AFM) phase.
To accommodate the magnetic structure,
we use a 2$\times$1$\times$1 supercell. \cite{wang07}
For the supercell we used, a
$1\times2\times4$ Monkhorst-Pack k-points mesh
converges very well the results.

\begin{table*}
\caption{Calculated phonon frequencies of ground state
structure $L$. The experiment vaules for Raman phonons are extracted
from Ref. \onlinecite{flores06}, whereas the IR phonons frequencies
are extracted
from Ref. \onlinecite{cao08}.
}
\begin{center}
\begin{tabular}{cccccccccccccccc}
\hline
\hline
\multicolumn{2}{c}{$B_{2u}$}&\multicolumn{2}{c}{$A_g$}
&\multicolumn{2}{c}{$B_{2g}$}&\multicolumn{2}{c}{$A_u$}
&\multicolumn{2}{c}{$B_{3u}$}&\multicolumn{2}{c}{$B_{1g}$}
&\multicolumn{2}{c}{$B_{1u}$}&\multicolumn{2}{c}{$B_{3g}$}\\ 

GGA &Exp.&GGA &Exp.&GGA &Exp.&GGA &Exp. &GGA &Exp.&GGA
&Exp.&GGA&Exp.&GGA &Exp.\\\hline

671.8  &680.0  &679.2  &695.0  &566.0  &       &614.1  &       &677.2  &713.0  &683.7  &675.0  &607.2  &       &557.2  &585.0 \\

628.4  &       &617.7  &625.0  &480.8  &510.0  &531.2  &       &620.1  &       &647.0  &       &530.0  &       &507.9  &495.0 \\

553.4  &       &615.0  &       &470.8  &       &499.5  &       &597.1  &       &588.3  &       &497.9  &       &471.7  &      \\

535.8  &       &531.5  &545.0  &452.3  &460.0  &429.1  &       &512.7  &519.0  &547.5  &540.0  &437.2  &       &450.4  &440.0 \\

473.8  &       &491.7  &500.0  &444.7  &       &402.3  &       &475.6  &       &521.2  &       &410.6  &403.0  &430.5  &      \\

438.0  &       &444.2  &460.0  &336.5  &       &287.5  &       &465.5  &       &470.1  &485.0  &296.7  &290.0  &324.7  &320.0 \\

412.1  &       &407.2  &420.0  &293.0  &305.0  &225.7  &       &378.3  &       &411.0  &420.0  &249.3  &       &289.8  &      \\

351.0  &       &336.9  &350.0  &273.6  &       &132.5  &       &351.5  &       &371.3  &330.0  &142.5  &       &264.5  &      \\

314.4  &       &311.1  &       &230.7  &220.0  &110.8  &       &317.0  &       &299.6  &       & -2.0  &       &241.5  &      \\

267.8  &267.0  &235.4  &       &209.6  &       &       &       &265.4  &       &242.6  &235.0  &       &       &189.8  &      \\

223.0  &       &224.2  &215.0  & 96.6  &       &       &       &216.3  &217.0  &220.2  &205.0  &       &       &103.6  &      \\

163.2  &       &136.4  &       &       &       &       &       &158.2  &       &171.9  &       &       &       &       &      \\

157.5  &       &108.4  &       &       &       &       &       &151.5  &       &146.1  &145.0  &       &       &       &      \\

 99.6  &100.0  &       &       &       &       &       &       &111.6  &       &       &       &       &       &       &      \\

\hline\hline
\end{tabular}
\label{tab:phonon-L}
\end{center}
\end{table*}

\section{Results and Discussion}

The fully relaxed DyMn$_2$O$_5$ is an AFM insulator, and
has same lattice structure
to that of TbMn$_2$O$_5$, \cite{alonso97,wang07}
both are of $Pb2_1m$ symmetry,
as Tb and Dy are isovalent.
The lattice structure is distorted from a $Pbam$
high ($H$) symmetry structure, due to the spin-lattice coupling.
\cite{wang07}
The calculated lattice constants
are $a$=7.270 \AA, $b$=8.518 \AA \, and $c$=5.600 \AA, respectively,
and are in very good agreement in experimental values \cite{blake05}
(7.285, 8.487 and 5.668 \AA\, respectively).
The lattice constants are
somewhat smaller than those of TbMn$_2$O$_5$. \cite{alonso97,wang07}
Like TbM$_2$O$_5$, \cite{chapon04, wang07}
DyMn$_2$O$_5$ has two energetically degenerate lattice (and magnetic) structures,
$L$ and $R$,
in which Mn$^{4+}$ form an
AFM square lattice in the $ab$ plane, whereas Mn$^{3+}$ couples to
Mn$^{4+}$ either antiferromagnetically via $J_4$ along $a$ axis or
with alternating sign via $J_3$ along $b$ axis.
Mn$^{3+}$ ions in two connected pyramids also couple
antiferromagnetically through $J_5$. Here, we adopt the notations
$J_3$, $J_4$ and $J_5$ from Ref. \onlinecite{chapon04}, and define
the $J_3$ to be the Mn$^{4+}$- Mn$^{3+}$ superexchange interaction
through pyramidal base corners, and $J_4$ the superexchange
interaction through the pyramidal apex.

We first calculate all zone-center optical
phonon frequencies of the
fully relaxed structure ($L$ or $R$) via
a frozen-phonon technique, \cite{wang07}
and the results are listed in
Table \ref{tab:phonon-L}.
Symmetry analysis \cite{wang08}
shows that
for the high symmetry ($Pbam$) structure $H$, phonons belong to 8
irreducible representations (irreps),
among them $B_{1u}$, $B_{2u}$ and $B_{3u}$ are IR active, where
$B_{1g}$, $B_{2g}$, $B_{3g}$ and $A_g$ are Raman active and
$A_u$ is silent.
However, in the low symmetry  ($Pb21m$) structure ($L$ or $R$),
the Raman and IR modes couple to each other, and regroup to
4 irreps, i.e.,
$A_1 =B_{2u} \oplus A_g$; $A_2 =B_{2g} \oplus A_u$;
$B_2=B_{3u} \oplus B_{1g}$;
$B_1=B_{1u} \oplus B_{3g}$.
The couplings between irreps. are very small, therefore
the results are given by their major symmetry characters.
As we see from Table \ref{tab:phonon-L}, the results are
in excellent agreement with experiments. \cite{flores06, cao08}
The calculated phonon frequencies are very close
to those of TbMn$_2$O$_5$, because Tb and Dy are isovalent
and the two materials have very similar lattice structures.
The inner products between the corresponding phonon modes \cite{he03}
of the two compounds are close to unity, suggesting that they have very
similar mode patterns.

Experimentally, there are several anomalies found for both Raman modes and IR
modes, near the magnetic phase transitions. For example,
anomalous phonon shift for the Raman modes
had been observed at T$^*$ $\sim$ 60-65 K, and near the N\'{e}el temperature
T$_{N}$. \cite{flores06}
The highest A$_g$ mode show a steep hardening upon cooling between
T$^*$ and T$_N$. \cite{flores06}
For the IR modes, it has been found
near 60K, the infrared active modes soften and several modes soften
substantially. \cite{cao08b}
The phonon modes are also found to be
very sensitive to the applied magnetic field. \cite{cao08}
The observed phonon anomalous phonon shift
strongly suggest that the spin-phonon coupling in this material is
very strong.

Theoretically, the spin-lattice coupling in this compound can be described via
an effect Hamiltonian, \cite{wang08}
\begin{widetext}
\begin{equation}
E(\{u_{\lambda} \}) =(E_0-\sum_{ij}J_{ij} {\bf S}_i\cdot {\bf S}_j)
-\sum_{ij}\sum_{\lambda} {\partial J_{ij} \over \partial
u_{\lambda}} u_{\lambda} {\bf S}_i\cdot {\bf S}_j
+({1 \over 2}\sum_{\lambda} m_{\lambda} \omega_{\lambda}^{2}u_{\lambda}^{2}
-\sum_{ij}\sum_{\lambda\rho}{\partial^{2}J_{ij} \over \partial u_{\lambda}
\partial u_{\rho}}u_{\lambda}u_{\rho}{\bf S}_i\cdot {\bf S}_j)\, ,
\label{eq:Heff}
\end{equation}
\end{widetext}
where $u_{\lambda}$ is the $\lambda$-th zone-center phonon,
and ${\bf S}_i$ is the magnetic moment of the $i$-th atom.
Here, we consider
only the magnetic moments of the Mn$^{3+}$ and Mn$^{4+}$ ions.
$J_{ij}$ is the exchange interaction between
two adjacent Mn ions.
$E_0$ is the total energy of high symmetry structure $H$
without spin-spin interactions, where
the ground state spin configuration is determined by the first term.
$\omega_{\lambda}$ is the frequencies of the $\lambda$-th phonon,
calculated in the high symmetry PM state.
The spin-lattice coupling have two effects:
(i) The
second term, which is linear in $u$, breaks the
inversion symmetry of the system, and drive
the structure to a low symmetry polarized state leading to coexistence of AFM
and ferroelectricity. The coupling strength is proportional to
${\partial J_{ij}/ \partial u_{\lambda}}$.
However, this term would not
change the phonon frequencies (at least to the first order approximation).
(ii) The phonon modes $\omega_{\lambda}$
(including frequencies and eigenvectors)
are modified in the presence of spin-phonon interaction, due to the
third term of Eq. (\ref{eq:Heff}),
where the spin-phonon coupling strength is
${\partial^{2}J_{ij}/ \partial u_{\lambda}\partial u_{\rho}}$.
Different spin configuration (SC) $\{ {\bf S}_i \}$ would therefore
have different phonon frequencies.

\begin{figure}
\centering
\includegraphics[width=2.8in]{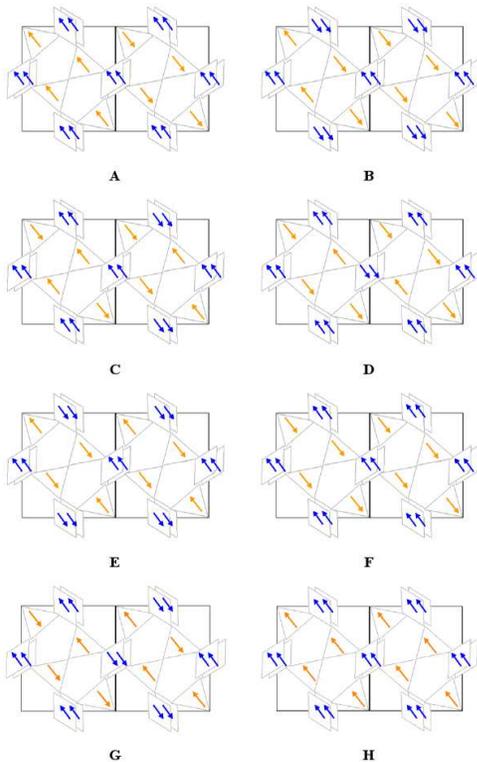}
\caption{(Color online) The spin configurations used in calculating phonon
  frequencies of different spin states.}
\label{fig:spins}
\end{figure}

\begin{table*}
\caption{The $A_g$ modes calculated in high symmetry structure $H$
of different spin configurations using GGA and GGA + U methods.}
\begin{center}
\begin{tabular} 
{p{1.5cm} p{1.5cm} p{1.5cm} p{1.5cm}p{1.5cm} p{1.5cm} p{1.5cm} p{1.5cm} p{1.5cm} }
\hline\hline
\mc{1}{c}{\it u}-GGA  &\mc{4}{c}{GGA} &\mc{4}{c} {GGA + U} \\
 & $\langle \omega \rangle_{dis} $ & $\langle \omega \rangle_{j4}$ & AFM 
&FM & $\langle \omega \rangle_{dis}$ & $\langle \omega \rangle_{j4}$ &AFM &FM\\
\hline

571.1  & 667.7 & 677.6 & 679.0 & 662.8 &   746.1& 759.9 & 759.0 &723.2\\
563.2  & 621.5 & 625.8 & 618.6 & 629.7 &   658.8& 661.3 & 659.4 &653.4\\
530.7  & 617.7 & 618.4 & 613.5 & 617.3 &   634.2& 635.3 & 631.0 &638.3\\
463.3  & 537.2 & 540.9 & 531.1 & 547.0 &   559.3& 561.1 & 556.6 &562.2\\
432.9  & 490.6 & 492.2 & 491.3 & 494.1 &   505.4& 506.3 & 506.0 &506.4\\
409.2  & 441.0 & 442.8 & 442.6 & 441.4 &   451.4& 452.7 & 452.6 &450.1\\
369.6  & 414.3 & 407.9 & 407.0 & 429.8 &   438.2& 436.6 & 436.0 &440.3\\
249.6  & 340.0 & 337.6 & 337.6 & 340.3 &   342.6& 342.3 & 341.6 &343.0\\
178.8  & 321.9 & 329.2 & 311.0 & 331.7 &   328.9& 331.9 & 323.2 &334.8\\
152.7  & 238.7 & 237.2 & 236.2 & 243.9 &   248.3& 248.0 & 247.5 &249.2\\
102.9  & 222.3 & 220.2 & 219.6 & 225.9 &   217.7& 216.9 & 216.6 &220.1\\
66.1   & 136.8 & 137.5 & 136.0 & 137.6 &   141.1& 141.6 & 140.9 &141.0\\
-174.2 & 110.0 & 110.0 & 109.3 & 111.2 &   113.2& 113.4 & 113.1 &113.5\\

\hline
\end{tabular}
\label{tab:phonon-Ag}
\end{center}
\end{table*}

\begin{table*}
\caption{The $B_{2u}$ modes calculated in high symmetry structure $H$
of different spin configurations using GGA and GGA + U methods.}
\begin{center}
\begin{tabular} 
{p{1.5cm} p{1.5cm} p{1.5cm} p{1.5cm}p{1.5cm} p{1.5cm} p{1.5cm} p{1.5cm} p{1.5cm} }
\hline\hline
\mc{1}{c}{\it u}-GGA  &\mc{4}{c}{GGA} &\mc{4}{c} {GGA + U} \\
 & $\langle \omega \rangle_{dis} $ & $\langle \omega \rangle_{j4}$ & AFM 
&FM & $\langle \omega \rangle_{dis}$ & $\langle \omega \rangle_{j4}$ &AFM &FM\\
\hline

586.4 & 660.2 & 673.3 & 671.7 & 649.6 & 723.7 & 736.6 & 735.9 &705.8\\
512.8 & 627.2 & 625.5 & 626.1 & 623.8 & 659.1 & 659.4 & 659.1 &657.9\\
507.2 & 551.6 & 552.2 & 553.4 & 552.3 & 578.5 & 578.7 & 579.4 &577.9\\
488.7 & 534.0 & 530.9 & 535.3 & 534.1 & 566.6 & 567.0 & 568.3 &564.6\\
431.2 & 474.5 & 473.5 & 475.7 & 476.4 & 495.5 & 494.5 & 495.6 &498.1\\
391.0 & 439.5 & 437.2 & 439.2 & 438.1 & 461.8 & 462.0 & 462.2 &459.4\\
381.3 & 409.7 & 410.0 & 411.7 & 406.0 & 419.9 & 420.2 & 420.9 &418.0\\
310.0 & 353.6 & 350.2 & 350.8 & 357.6 & 365.7 & 364.9 & 365.2 &366.1\\
274.3 & 315.0 & 314.3 & 315.2 & 314.6 & 316.2 & 315.8 & 316.3 &316.6\\
257.7 & 272.7 & 266.0 & 267.9 & 280.1 & 283.8 & 282.5 & 282.5 &285.5\\
211.1 & 222.2 & 221.8 & 223.4 & 220.8 & 211.7 & 212.0 & 212.5 &212.7\\
142.6 & 162.0 & 162.2 & 162.0 & 161.4 & 165.6 & 165.7 & 165.5 &165.7\\
139.2 & 156.8 & 156.9 & 157.0 & 157.3 & 163.3 & 163.5 & 163.7 &162.7\\
93.9  & 98.4 & 98.6 & 99.2 & 96.8  &  94.0 &  94.3 &  94.6    &93.4\\

\hline
\end{tabular}
\label{tab:phonon-B2u}
\end{center}
\end{table*}

To investigate how the SCs
change the phonons in DyMn$_2$O$_5$,
we calculate the phonons frequencies of different
SCs including the high
temperature PM state, the AFM state,
and the ferromagnetic (FM) state.
The spin-lattice coupling strength $J''$
can be extracted by
comparing the force-constant matrices calculated from
these different spin states. \cite{fennie06}
[see Eq. (\ref{eq:Heff})].
In this work,
we focus on the B$_{2u}$ and A$_g$ modes, whereas other
modes can be studied in similar way.
To simplify the discussion,
all following calculations are done in the
high symmetry structure $H$, constructed via symmetrizing
structures $L$ and $R$ according to the $Pbam$ symmetry. \cite{wang07}
In reality, the lattice constants
would be somewhat different at different SCs.
In the present calculations,
this effect is ignored and
the lattice constants are fixed in the calculations.

The phonon frequencies of PM, AFM (SC G in Fig. \ref{fig:spins}) 
and FM (SC H in Fig. \ref{fig:spins})  states are compared in  
Table~\ref{tab:phonon-Ag} for the A$_g$ modes and in
Table \ref{tab:phonon-B2u} for
the B$_{2u}$ modes.
One may attempt to calculate the phonon frequencies of the PM state
using spin {\it unpolarized} GGA, which is listed in the
column under {\it u}-GGA.
A quick look reveals that the phonon frequencies
in this column are significantly lower than those of
AFM and FM state.
Especially, as shown in Table~\ref{tab:phonon-Ag},
the A$_g$ irrep has a soft mode,
causing a Jahn-Teller distortion. \cite{kittel_book}
The phonon frequencies calculated from $u$-GGA is not a good
approximation for the PM state, because
the Mn ions have local magnetic moments even in the PM phase,
although their directions are distributed randomly.
Alternatively, we calculate $\langle \omega \rangle_{dis}$,
which are the averaged phonon frequencies of several fully
disordered SCs ($A$, $B$, $C$ in Fig. \ref{fig:spins} ) 
so all exchange interactions are canceled out.
The averaged phonon frequencies
$\langle \omega \rangle_{dis}$ are close to those
of the AFM and FM states without soft phonons.

However, the spins are fully disordered only at very high temperature.
At lower temperature, especially close to the magnetic phase transition, the
spins are somehow correlated, and develop short range ordering.
Since $J_4$ is the largest among all the exchange interactions,
\cite{wang08}
to compare with experiments,
we consider the ordering of $J_4$ interactions. We calculate
$\langle \omega \rangle_{J_4}$, which are the averaged phonon frequencies
of three SCs (D, E, F in Fig. \ref{fig:spins}) 
in which $J_4$ are fixed in the AFM
configuration, whereas all other exchange interactions are canceled out.
As we see from Table~\ref{tab:phonon-Ag} and Table~\ref{tab:phonon-B2u},
the frequencies of most modes in different SCs differ
by about 3 - 5 cm$^{-1}$, which are of similar magnitude to 
the phonon shifts in the magnetic field experiment. \cite{cao08}
However, the high frequency B$_{2u}$
(671 cm$^{-1}$) and A$_g$  (679 cm$^{-1}$) modes 
change dramatically
($\sim$ 20 cm$^{-1}$)
for different SCs.
Especially, $\langle \omega \rangle_{dis}$
and $\langle \omega \rangle_{J_4}$ are different
by about 10 and 13 cm$^{-1}$ for the $A_g$ and $B_{2u}$ 
modes, respectively, though both belong to the PM phase.
Interestingly, the two modes frequencies of $\langle \omega \rangle_{J_4}$
are close to the those in
AFM state, and  $\langle \omega \rangle_{dis} $ have the mean values 
of those of AFM and FM states.
It is usually believed that GGA overestimates the exchange
interactions, whereas including the on-site Coulomb correlation 
may improve this problem. To see how including the  
on-site Coulomb correlation 
will change the spin-phonon coupling in
DyMn$_2$O$_5$, we have carried out GGA + U calculations of the phonons
in different SCs. The on-site Coulomb $U$ of Mn ions has been taken
to be a reasonable value 4.0 eV, 
whereas the exchange parameter $j$= 0.8 eV is used.
The results are also listed in Table \ref{tab:phonon-Ag} and Table
  \ref{tab:phonon-B2u} for the A$_g$ and
B$_{2u}$ modes respectively. As we see, GGA + U significantly 
overestimates the frequencies of the high frequency modes 
when compared to experiments. However, the frequency
  difference between different SCs are almost remain the same to those
  without $U$, suggesting that 
the spin-phonon coupling constants do not change much
by including the  
on-site Coulomb correlation. In the following discussion, we therefore 
focus only the GGA results.

\begin{figure}
\centering
\includegraphics[width=2.65in]{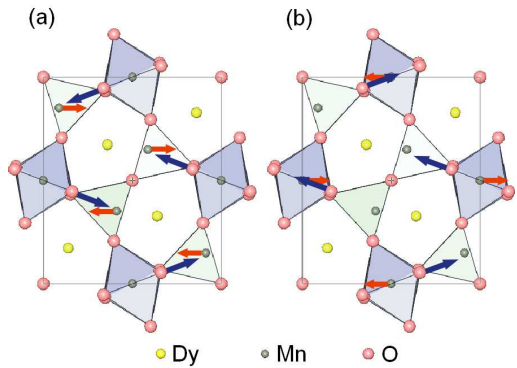}
\caption{(Color online) The mode pattern of
the highest (a) A$_g$ mode and (b) B$_{2u}$
mode.}
\label{fig:mode}
\end{figure}

To understand the results, we analyze the vibrational 
pattern of the two modes,
shown in Fig. \ref{fig:mode}.
For the high-frequency A$_g$ mode, about 90$\%$ of the
vibration is associated with O$_3$ (the oxygens that
connect Mn$^{3+}$ and Mn$^{4+}$ along the $a$ axis,
and convey $J_4$ interaction). The O$_3$ atoms vibrate
in the $ab$ plane, with an angle of 21.4$^\circ$ to $a$ axis.
This mode also has small components of Mn$^{3+}$ 
vibrating in the $a$-axis,
and Mn$^{4+}$ motion in the $c$-axis. Therefore $J_5$ may
also affect the A$_g$ mode (but considerably smaller than $J_4$).
In the high-frequency B$_{2u}$ mode, more than 70$\%$ of
the contribution comes from O$_3$, which also vibrate in
the $ab$ plane, with an angle of 19.9$^\circ$ to $a$ axis.
The rest contribution includes the motion of Mn$^{4+}$ along $a$ axis.
If the spin exchange interaction is local, the motion the O$_3$ atoms
would only tune the $J_4$ interactions. 
Therefore the $J_4$ interaction plays
an essential role of the spin-phonon coupling in these two modes.
Now it is easy to understand the phonon shifts of the two modes in
different SCs.
According to
Eq. (\ref{eq:Heff}), the frequency shifts of the
highest A$_g$ and B$_{2u}$ modes are
$\langle {\bf S}_3 \cdot {\bf S}_4 \rangle \,
{\partial^{2}J_{4} / \partial u_{\lambda}^2} $, where ${\bf S}_3$,
${\bf S}_4$ are the spin vectors
of Mn$^{3+}$ and Mn$^{4+}$ ions associated with $J_4$, respectively. 
For the fully disordered SC, 
$\langle {\bf S}_3 \cdot {\bf S}_4 \rangle$=0.
In the $J_4$ short range orderd state
and in AFM state 
$\langle {\bf S}_3 \cdot {\bf S}_4 \rangle$=-$S_3 S_4$,
whereas in FM state,
$\langle {\bf S}_3 \cdot {\bf S}_4 \rangle$=$S_3 S_4$.
$S_3$=2.3 $\mu_B$ and $S_4$=1.64 $\mu_B$, 
are the local magnetic moments.
By comparing the force-constant
matrices of different SCs,
we estimate ${\partial^{2}J_{4} / \partial u_{\lambda}^2} \sim$ 0.120
meV/$(\rm{\AA}\cdot \mu_B)^2$ for the two high frequency modes.
As discussed in previous
section, this value remain
unchanged when including the on-site Coulomb interactions.

Experimentally, such large phonon hardening
due to $J_4$ ordering was not directly observed
during the temperature dependent measurement \cite{flores06,cao08b}
for two reasons. First,
the $J_4$ ordering develop gradually 
with decreasing of the temperature.
%
A Monte Carlo simulation \cite{caok_unpub} 
show that the average of
$\langle {\bf S}_3 \cdot {\bf S}_4 \rangle$
increases from 80\% to 90\% of its maximum value
when temperature lower from 1.5 T$_N$
to T$_N$.
Second, the phonon frequencies hardening due to $J_4$ ordering is
accompanied by the anharmonic effects, \cite{flores06,cao08b}
 and it is hard to isolate
the  anharmonic effects and spin-phonon
effects in experiments.
The lower frequencies modes involve collective motion of many atoms.
Therefore, the phonon frequency shifts due to spin-phonon coupling
are a consequence of the
competition of many $J''$s, and are much smaller than the
high frequency modes.

Below N\'{e}el temperature, T$_N$, the material is in
the long range ordered AFM state.
We then compare the phonon frequencies of short range
ordered state $\langle \omega \rangle_{J_4}$ to those of
AFM state.
As we see, while the highest A$_g$ mode hardens by 2 cm$^{-1}$, and
the highest B$_{2u}$ mode softs by  2 cm$^{-1}$,
consistent with experiments.\cite{flores06,cao08b}
The frequencies difference between 
$\langle \omega \rangle_{J_4}$ and AFM might 
come from the $J_5$ interactions.

\section{summary}

To summarize, we have investigated the spin-phonon
coupling in a spin frustrated DyMn$_2$O$_5$ system
via first-principles calculations.
We compare the phonon modes calculated from different
spin configurations.
The results show that the short range spin ordering
can dramatically change the phonon
frequencies in this compound, and might be responsible for the
observed phonon anomalies near and above the
magnetic phase transitions.

We would like to thank J. L. Musfeldt and J. Cao 
for communicating results prior to publication.
L.H. acknowledges the support from the Chinese National
Fundamental Research Program 2006CB921900, the Innovation
funds and ``Hundreds of Talents'' program from Chinese Academy of
Sciences, and National Natural Science Foundation of China.


\end{document}